\documentclass[twocolumn,amsmath,amssymb,floatfix,pra,aps,showpacs,articles,nofootinbib,groupaddresses,superscriptaddress]{revtex4}

\usepackage{amsmath}
\usepackage{amsthm}
\usepackage{amsfonts}
\usepackage{amssymb}
\usepackage{graphicx,graphics,times,color}
\usepackage{bbm}
\usepackage{subfigure}

\usepackage{color}

\begin{document}

\title{Optimal feedback control of two-qubit entanglement in dissipative environments}

\author{Morteza Rafiee}
\email{m.rafiee178@gmail.com}
\affiliation{Department of Physics, Shahrood University of Technology , 3619995161 Shahrood, Iran}

\author{Alireza Nourmandipour}
\email{anoormandip@stu.yazd.ac.ir}
\affiliation{Atomic and Molecular Group, Faculty of Physics, Yazd University, Yazd 89195-741, Iran}

\author{Stefano Mancini}
\email{stefano.mancini@unicam.it}
\affiliation{School of Science \& Technology, University of Camerino, I-62032 Camerino, Italy}
\affiliation{INFN-Sezione di Perugia, Via A. Pascoli, I-06123 Perugia, Italy}

\begin{abstract}
We study the correction of errors intervening in two-qubit dissipating into their own environments. This is done by resorting to local feedback actions with the aim of preserving as much as possible the initial amount of entanglement. Optimal control is found by first gaining insights from the subsystem purity and then by numerical analysis on the concurrence.
This is tantamount to a double optimization, on the actuation and on the measurement
precesses. Repeated feedback action is also investigated, thus paving the way for a continuous time formulation and solution of the problem.
\end{abstract}


\maketitle

\section{Introduction}

The feature of quantum mechanics which most distinguishes it from classical mechanics is the coherent superposition of distinct physical states, usually referred to as quantum coherence. It embraces also entanglement, i.e. non-local quantum correlations arising in composite systems \cite{HHHH2009}.
Quantum coherence results rather fragile against environment effects and this fact has boosted the development of a quantum control theory \cite{MMW2005}.
Just like the classical one, quantum control theory includes open-loop control
and closed-loop control according to the principle of controllers design \cite{Rabitz2009}.
Feedback is a paradigm of closed loop control, in that it involves gathering information about the system state and then according to that actuate a corrective action on its dynamics.
It has been shown that quantum feedback is superior to open-loop control in dealing with uncertainties in initial states \cite{Dong2010}. Moreover, it has been proven that it works better than open-loop control when it aims at restoring quantum coherence \cite{Qi2010}.

In the presence of feedback, suitable quantum operations are added to the bare dynamical map (resulting from the environment action) of a quantum system. These quantum operations should be determined according to the desired target state. This is like to say that one optimizes the \emph{actuation}. Besides, it is known that there is a correspondence between measurement on the environment and the representation of the map \cite{Kraus1991}. Therefore, it is clear that one has to optimize the \emph{measurement} overall possible representations of the map
in order to to extract the maximum information with the minimum disturbance.
Altogether, it can be said that feedback implies in the quantum realm a double optimization, over the measurement and over the actuation process \cite{Jacobs06}.
This makes designing the optimal feedback control a daunting task for quantum systems, especially composite ones and hence entanglement control (we refer here to \emph{local} control, i.e. measurement and actuation are both local operations).
In linear bosonic systems the pursued strategy was to steer a system towards a stationary state entangled as much as possible \cite{Mancini2007}.
Dealing with the inherent nonlinearity of qubits makes this strategy very challenging and no progresses have been made since the seminal work of Ref.\cite{Mancini2005}.

Hence, we shall consider here a feedback control whose aim is to preserve as much as possible an initial maximally entangled states for two-qubit dissipating into their own environments. Actually we shall employ maps and corrective actions much in the spirit of \cite{Memarzadeh2011NJPHYS}, without analyzing continuous time evolution.
Optimal control is found by first gaining insights from the subsystem purity and then by numerical analysis on the concurrence.
Repeated feedback action is also investigated, thus paving the way for a continuous time formulation and solution of the problem.

The layout of the paper is as follows. We start by introducing the model in Sec.\ref{sec:model}.
Then we discuss the feedback action in Sec.\ref{sec:FB} and subsequently address its optimality in Sec.\ref{sec:opt}.
Sec.\ref{sec:repeat} is devoted to repeated applications of the dynamical map.
Finally, Sec.\ref{sec:conclu} is for conclusion.


\section{The Model}\label{sec:model}

We consider two qubits (distinguished whenever necessary by labels $A$ and $B$) undergoing the effect of local amplitude damping, so that their initial state $\rho$ changes according to the following quantum channel map
\begin{equation}\label{dynnofbnew}
\rho\mapsto \rho'=\sum_{j=1}^4 K_j \rho K_j^\dag,
\end{equation}
where
\begin{equation}
\begin{aligned}
K_1&=E_1\otimes E_1, \\
K_2&=E_1\otimes E_2, \\
K_3&=E_2\otimes E_1,\\
K_4&=E_2\otimes E_2,
\end{aligned}
\label{eq:krausnew}
\end{equation}
are the Kraus operators (satisfying $\sum_{i=1}^{4}K_i^{\dagger}K_i=I$) constructed from those of local (single qubit) amplitude damping channels
\begin{equation}\label{eq:amplitudedamping}
\begin{aligned}
E_1&=\left(\sqrt{\eta}|1\rangle\langle 1|+|0\rangle\langle 0|\right),\\
E_2&=\left(\sqrt{1-\eta}|0\rangle\langle 1|\right).
\end{aligned}
\end{equation}
Here $|0\rangle$ (resp. $|1\rangle$) is the ground (resp. excited) qubit state and
$\eta\in\left[ 0,1\right] $ is the single qubit damping rate.

The map \eqref{dynnofbnew} implies the probability for each qubit of losing independently the excitation into its own environment.

Suppose that the two qubits are initially prepared in a maximally entangled states,
  e.g. $\rho=\vert \Phi \rangle   \langle \Phi \vert$ with
\begin{align}
    \vert \Phi \rangle &:= \frac{\vert 00 \rangle + \vert 11 \rangle}{\sqrt{2}}.
\end{align}
In the computational basis  $\mathfrak{B}:=\{ |11\rangle, |10\rangle, |01\rangle, |00\rangle \}$,
it has the following matrix representation:
\begin{equation}
\begin{aligned}
\rho=\dfrac{1}{2}
&\begin{pmatrix}
1 & 0 & 0 & 1 \\
0 & 0 & 0 & 0 \\
0 & 0 & 0 & 0 \\
1 & 0 & 0 & 1 \\
 \end{pmatrix}.
 \end{aligned}
 \label{eq:initalstate}
\end{equation}
From here on we assume the freedom to perform local operations (and eventually classical communication), i.e. they are costless. Hence the above assumption of the initial state is equivalent to any other maximally entangled state.

In the computational basis  $\mathfrak{B}$, the state $\rho^\prime$ resulting from Eq.\eqref{dynnofbnew} reads
\begin{equation}
\begin{aligned}
\rho^\prime=\dfrac{1}{2}
&\begin{pmatrix}
\eta^2 & 0 & 0 & \eta \\
0 & \eta(1-\eta) & 0 & 0 \\
0 & 0 & \eta(1-\eta) & 0 \\
\eta & 0 & 0 & 2+\eta(\eta-2) \\
 \end{pmatrix}.
 \end{aligned}
 \label{dynnofbnewrho}
\end{equation}
Now consider the subsystem purity
\begin{equation}
{\cal P}(\rho):={\rm Tr}(\rho_A^2), \quad \rho_A:={\rm Tr}_B\rho,
\end{equation}
as measure of entanglement.
Although it is only valid for pure states $\rho$, it can give us some insights also for mixed states entanglement.
Thanks to  \eqref{dynnofbnewrho} it is straightforward to show that
\begin{align}\label{p1}
{\cal P}(\rho^\prime)=\frac{1}{2}\big(2-2\eta+\eta^2\big).
\end{align}
The minimum $1/2$ is achieved for $\eta=1$, i.e. when the channel \eqref{dynnofbnew} reduces to the identity map.

\begin{figure}[h!] \centering
	\includegraphics[width=8cm,height=6cm,angle=0]{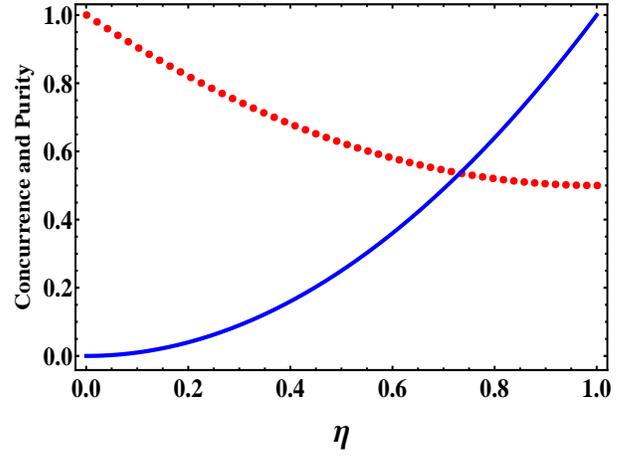}
	\caption{(Color online) Concurrence (solid blue line) and subsystem purity (dot red line) versus $\eta$ for the state $\rho'$.}
	\label{fig1}
\end{figure}

A faithful measure of entanglement is the concurrence defined as \cite{Wootters98}
\begin{equation}
C(\rho):=\max\{0,\lambda_1-\lambda_2-\lambda_3-\lambda_4\},
\end{equation}
where $\lambda_i$�s are, in decreasing order, the nonnegative square
roots of the moduli of the eigenvalues of
$\rho (\sigma_A^y\otimes \sigma_B^y)\rho^*(\sigma_A^y\otimes \sigma_B^y)$  with
$\rho^*$ denoting the complex conjugate of $\rho$.
Using \eqref{dynnofbnewrho}  we can show that
\begin{equation}
C(\rho^\prime)=\eta^2\sqrt{2+\eta^2-2\eta}-\frac{1}{2}\eta^2 (1-\eta^2).
\end{equation}
Fig. \ref{fig1} illustrates the subsystem purity as well as concurrence resulting from state  \eqref{dynnofbnew} as a function of $\eta$. We can see that they behave opposite one to another.
Hence we can argue that parameters minimizing the subsystem purity would also maximizing the concurrence.

\section{Feedback Action}\label{sec:FB}

The map in Eq.\eqref{dynnofbnew} can be regarded as the effect of a measurement process described by a
Probability Operator Valued Measure (POVM) whose elements are $\{K_j\}_j$ and whose outcomes
are labelled by the values of $j$ \cite{Kraus1991}.
Notice that the elements $\{K_j\}_j$ are local, hence
we consider \emph{local feedback} actions
$U_j\in$ SU(2)$\times$SU(2) to be applied in correspondence of the outcomes $j$s.
That is, in the presence of feedback the dynamical map \eqref{dynnofbnew} changes into
\begin{equation}\label{dynfbnew}
\rho\mapsto \rho''=\sum_{j=1}^4 \left( U_jK_j\right) \rho \left(U_jK_j\right)^\dag.
\end{equation}
Due to the symmetry of the initial state $\rho=   \vert \Phi \rangle   \langle \Phi \vert$ and of the action of the dissipative map, the unitary operators $U_j$s can be taken as:
\begin{equation}
\begin{aligned}\label{Ufb}
U_1&=u\otimes u,\\
U_2&=u\otimes v,\\
U_3&=v\otimes u,\\
U_4&=v\otimes v,
\end{aligned}
\end{equation}
where
\begin{equation}
\begin{aligned}\label{eq:uv}
u&=\begin{pmatrix}
   e^{-i(\alpha_u+\gamma_u)/2}\cos(\beta_u/2) & -e^{-i(\alpha_u-\gamma_u)/2}\sin(\beta_u/2) \\
   e^{i(\alpha_u-\gamma_u)/2}\sin(\beta_u/2) &  e^{i(\alpha_u+\gamma_u)/2}\cos(\beta_u/2)
   \end{pmatrix},\\ \\
v&= \begin{pmatrix}
   e^{-i(\alpha_v+\gamma_v)/2}\cos(\beta_v/2) & -e^{-i(\alpha_v-\gamma_v)/2}\sin(\beta_v/2) \\
   e^{i(\alpha_v-\gamma_v)/2}\sin(\beta_v/2) &  e^{i(\alpha_v+\gamma_v)/2}\cos(\beta_v/2)
   \end{pmatrix},
\end{aligned}
\end{equation}
are generic elements of SU(2) with $\alpha_\bullet$, $\beta_\bullet$,
$\gamma_\bullet$ the Euler angles.

This model makes fully sense because now once we are given an entangled state the feedback operations are completely local and the aim is to restore as much as possible entanglement (degraded by the local dissipation).
So the goal is to find the Euler angles that maximizes the amount of entanglement of $\rho''$.

\begin{figure}
 \centering
	\includegraphics[width=8cm,height=6cm,angle=0]{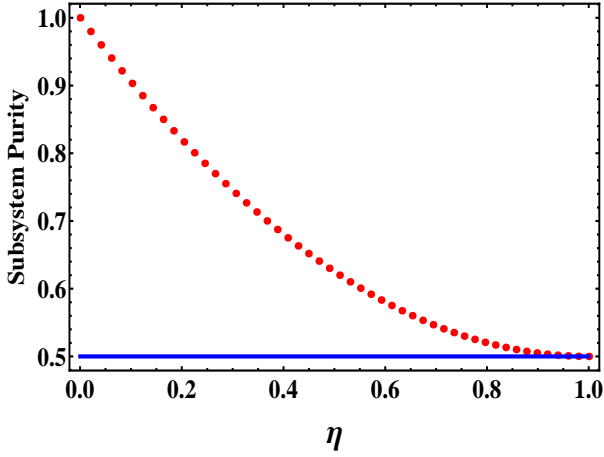}
	\caption{(Color online) Subsystem purity versus $\eta$ without feedback action (dot red line) and in the presence of feedback action with $\theta=-\phi=\pi$ (solid blue line). }
	\label{fig2}
\end{figure}

\begin{figure}
 \centering
	\includegraphics[width=8cm,height=6cm,angle=0]{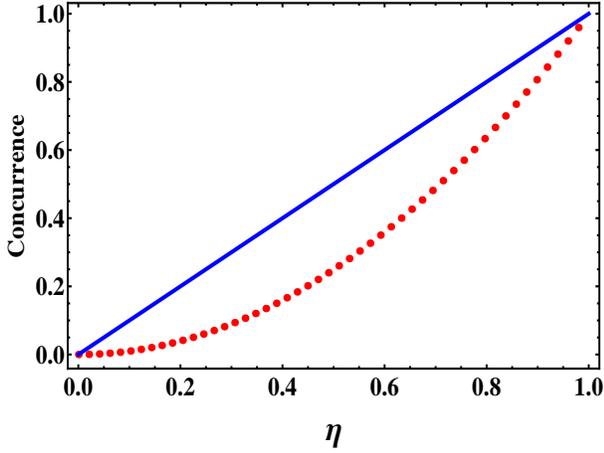}
	\caption{(Color online) Concurrence versus $\eta$ without feedback action (dot red line) and in the presence of feedback action with $\theta=-\phi=\pi$ (solid blue line).}
	\label{fig3}
\end{figure}

Applying \eqref{dynfbnew}, with \eqref{Ufb} and \eqref{eq:uv}, to
$\rho=   \vert \Phi\rangle   \langle \Phi \vert $ gives
$\rho''$ whose matrix elements in the basis $\mathfrak{B}$ are:
\begin{align}
\left[\rho''\right]_{11} &= \frac{1}{8} \bigg\{\left(1-\eta^2\right)\left( 1+ \cos\beta_v\right)^2+\left( 1+ \cos\beta_u\right)^2 \nonumber\\
&+8\eta(1-\eta)\cos ^2\left(\beta_v/2\right)\sin ^2\left(\beta_u/2\right) +4\eta^2\sin ^4\left(\beta_u/2\right) \nonumber\\
&+4\eta\left( 1+ \cos\beta_u\right)\cos (2 \gamma_u)\sin ^2\left(\beta_u/2\right)\bigg\}, \label{eq:r11} \\
\nonumber \\
\left[\rho''\right]_{12}&=\frac{1}{8} \bigg\{e^{-i \alpha_v}(1-\eta) \sin \beta_v \big[ (1-\eta) \cos \beta_v+1-\eta\cos\beta_u
\big] \nonumber \\
  &+e^{-i \alpha_u}\sin\beta_u\Big[(1-\eta)(1-\eta \cos \beta_v) +2 i \eta  \sin (2\gamma_u)  \nonumber\\
  &+\cos \beta_u\left(1+\eta ^2-2 \eta  \cos (2 \gamma_u)\right) \Big] \bigg\}, \label{eq:r12}
\end{align}

\begin{align}
\left[\rho''\right]_{13}&= \left[\rho''\right]_{12}, \\ \nonumber\\
\left[\rho''\right]_{14}&=\frac{1}{8}e^{-2 i (\alpha_u+\gamma_u)}\bigg\{ \eta(1+ \cos \beta_u)^2+4\eta e^{4 i\gamma_u}\sin ^2\left(\beta_u/2\right)\nonumber \\
&+ 2e^{2 i\gamma_u}(1+\eta^2)(1+\cos \beta_u) \sin ^2\left(\beta_u/2\right) \nonumber \\
&+ 2e^{2 i (\alpha_u-\alpha_v+\gamma_u)}(1-\eta)^2(1+\cos \beta_v) \sin ^2\left(\beta_v/2\right) \nonumber \\
&-2e^{ i (\alpha_u-\alpha_v+2\gamma_u)}\eta(1-\eta)\sin \beta_u\sin \beta_v
\bigg\},
\end{align}

\begin{align}
\left[\rho''\right]_{22}&= \frac{1}{8} \bigg\{4 (1-\eta) \eta  \cos ^2\left(\beta_u/2\right)
   \cos ^2\left(\beta_v/2\right) \nonumber\\
 &+ \left(1+\eta ^2-2 \eta  \cos (2 \gamma_u)\right)\sin ^2\beta_u  \nonumber \\
 &+2 (1-\eta) \big[1-\eta
    \cos\beta_u+(1-\eta) \cos \beta_v\big]\sin ^2\left(\beta_v/2\right) \bigg\}, \label{eq:r22}
\\ \nonumber \\
\left[\rho''\right]_{23}&=\frac{1}{8}\bigg\{\left(1+\eta ^2-2 \eta  \cos (2 \gamma_u)\right)\sin ^2\beta_u\nonumber\\
&-(1-\eta)\sin\beta_v \Big[2\eta\cos(\alpha_u-\alpha_v)\sin \beta_u \nonumber\\
&-(1-\eta)\sin\beta_v \Big]\bigg\}, \label{eq:r23}
\end{align}

\begin{align}
\left[\rho''\right]_{24}&=\frac{1}{8} \bigg\{
e^{-i \alpha_u}\sin\beta_u\Big[(1-\eta)(1+\eta \cos \beta_v) \nonumber\\
&-\cos \beta_u\left(1+\eta ^2-2 \eta  \cos (2 \gamma_u)\right) -2 i \eta  \sin (2\gamma_u) \Big] \nonumber\\
&+ e^{-i \alpha_v}(1-\eta) \big[1+\eta\cos\beta_u-(1-\eta) \cos \beta_v\big]\sin \beta_v
  \bigg\}, \label{eq:r24}\\
  \nonumber\\
\left[\rho''\right]_{33}&= \left[\rho''\right]_{22},
\end{align}

\begin{align}
\left[\rho''\right]_{34}&= \left[\rho''\right]_{24}, \\ \nonumber\\
\left[\rho''\right]_{44}&=\frac{1}{8} \bigg\{ \eta^2\left( 1+ \cos\beta_u\right)^2+4(1-\eta)^2\sin^4\left(\beta_v/2\right) \nonumber\\
&+4\sin^4\left(\beta_u/2\right)+8 (1-\eta) \eta  \cos ^2\left(\beta_u/2\right)
    \sin ^2\left(\beta_v/2\right) \nonumber\\
&+4\eta(1+\cos\beta_u)\cos(2\gamma_u)\sin ^2\left(\beta_u/2\right)\bigg\}.
 \label{eq:r44}
\end{align}
The subsystem purity for the state $\rho^{\prime\prime}$ reads
\begin{align}\label{p2}
{\cal P}(\rho'')=&\frac{1}{4} \bigg\{3- \eta (2-\eta) +\frac{1}{2} (1-\eta )^2 \nonumber\\
&\times\bigg[  (1-\cos \xi ) \cos \theta+(1+\cos \xi) \cos \phi \bigg]\bigg\},
\end{align}
where,
\begin{equation}
\begin{aligned}
\xi&:=\alpha_u-\alpha_v,\\
\theta&:=\beta_u + \beta_v, \\
\phi&:=\beta_u- \beta_v.
\end{aligned}
\end{equation}
Taking the partial derivatives of \eqref{p2} with respect to $\xi,$ $ \theta$ and $\phi$ and setting them equal to zero, we arrive at the following equations:
\begin{equation}
\begin{aligned}\label{derivative}
(\cos \xi -1) \sin \theta  &=0, \\
(\cos \theta -\cos \phi ) \sin \xi  &=0, \\
(\cos \xi +1) \sin \phi &=0.
\end{aligned}
\end{equation}
They have a set of solutions
\begin{equation}
\begin{aligned}
&\{\theta = 0,\xi = -\pi \}, \\
&\{\theta = 0,\xi = \pi \},  \\
&\{\theta = 0,\phi = 0\},  \\
&\{\phi = 0,\xi = 0\},
\end{aligned}
\end{equation}
which leads to the same amount of $\cal P$ without feedback.
The other set of solutions of \eqref{derivative}
\begin{equation}
\begin{aligned}\label{optsets}
&\{\theta = -\pi ,\phi = -\pi \}, \\
&\{\theta = -\pi ,\phi = \pi \},  \\
&\{\theta = \pi,\phi = -\pi \},   \\
&\{\theta = \pi ,\phi = \pi \},
\end{aligned}
\end{equation}
leads to constant subsystem purity equal to $1/2$ (minimum obtainable value) for any value of $\eta$ (and arbitrary value of $\xi$).
All the values in \eqref{optsets} give the following density operator
 \begin{equation}
\begin{aligned}
\rho''=\frac{1}{2}
&\begin{pmatrix}
1 & 0 & 0 & \eta e^{-2 i (\alpha_u+\gamma_u)} \\
0 & 0 & 0 &0 \\
0 & 0 & 0 & 0 \\
\eta e^{2 i (\alpha_u+\gamma_u)} & 0 & 0 & 1
 \end{pmatrix},
 \end{aligned}
 \label{rhoppopt}
\end{equation}
whose concurrence results
\begin{equation}\label{eq:Cfb}
C(\rho'')=\eta.
\end{equation}
The results for the subsystem purity and concurrence are displayed in Figs. \ref{fig2} and \ref{fig3}. They show that the behaviour of purity and concurrence versus $\eta$ are consistent.

\section{Optimality of Feedback Action}\label{sec:opt}

It is known that the same quantum channel can have many (actually infinite many) Kraus decompositions
and each one can be interpreted as a given measurement performed on the environment to gain information about the system \cite{Kraus1991}.
Hence, in this Section, we will check the optimality of feedback action on unitarily equivalent Kraus representation of map \eqref{dynnofbnew}. To this end, first notice that the Kraus representation provided in \eqref{eq:krausnew} is canonical,
i.e. ${\rm Tr}\left(K_i^\dag K_j\right)\propto \delta_{ij}$. Then, restricting to canonical Kraus operators,
we should consider new Kraus operators
\begin{equation}\label{eq:tildeKj}
\widetilde{K}_i=\sum_{j=1}^4V_{ij}K_j,
\end{equation}
obtainable by linear combination of the old ones through a unitary matrix $V=V_A\otimes V_B$, in which
$V_A, V_B\in$ SU(2) and similarly to \eqref{eq:uv} can be parametrized as
\begin{equation}
\begin{aligned}
V_A=
&\begin{pmatrix}
\alpha & \beta \\
-\beta^{*}  & \alpha^{*}
 \end{pmatrix},\\ \\
V_B=
&\begin{pmatrix}
\alpha' & \beta' \\
-\beta'^{*}  & \alpha'^{*}
 \end{pmatrix},
 \end{aligned}
\end{equation}
with $|\alpha|^2+|\beta|^2=|\alpha'|^2+|\beta|^2=1$.
Explicitly we have
\begin{equation}\label{newKraus}
\begin{aligned}
\widetilde{K}_1&=\alpha\alpha' K_1+\alpha  \beta'  K_2+\alpha'  \beta  K_3+\beta\beta' K_4, \\
\widetilde{K}_2&=-\alpha  \beta'^{*}K_1+\alpha\alpha'^{*} K_2-\beta\beta'^{*} K_3+\beta \alpha'^{*} K_4,  \\
\widetilde{K}_3&=-\alpha'  \beta ^*K_1-\beta'\beta^* K_2+\alpha'\alpha^* K_3+\beta'  \alpha ^* K_4,  \\
\widetilde{K}_4&=  \beta^*\beta'^{*} K_1 -\beta ^*\alpha'^{*}K_2-\alpha^* \beta'^{*}K_3+\alpha ^*\alpha'^{*}K_4.
\end{aligned}
\end{equation}
This means we can now describe the dynamics of the density matrix
in the presence of feedback as
\begin{equation}\label{eq:rhotrans}
\rho\mapsto \widetilde{\rho}^{\;\prime\prime}=\sum_{j=1}^4 \left( U_j \widetilde{K}_j\right) \rho \left(U_j\widetilde{K}_j\right)^\dag.
\end{equation}
The expression of $\widetilde{\rho}^{\prime\prime}$ is too cumbersome to be reported here. However, computing its subsystem purity, the surprising aspect is that it becomes function of only $\left\lbrace \alpha,\beta\right\rbrace $.
Actually it reads
\begin{equation}\label{purityV}
 {\cal P}(\widetilde{\rho}^{\;\prime\prime})= \frac{1}{8}\Big(4+P_1^2+\left| P_2\right| ^2\Big),
  \end{equation}
in which
\begin{align}\label{P1}
P_1&:=(1-\eta ) \big(\cos\beta_u+\cos\beta_v\big)  \nonumber\\
&+2 \sqrt{(1-\eta ) \eta } \sin\beta_u\Big(\sin\gamma_u\Im\left(\alpha  \beta
   ^*\right)-\cos\gamma_u\Re\left(\alpha
   \beta ^*\right)\Big) \nonumber \\
&-2 \sqrt{(1-\eta ) \eta } \sin\beta_v
      \Big(\sin\gamma_v\Im\left(\alpha  \beta
      ^*\right)-\cos\gamma_v\Re\left(\alpha
      \beta ^*\right)\Big), \\
\nonumber   \\
P_2&:= (1-\eta) \sin\beta_v+(1-\eta) e^{-i(\alpha_u-\alpha_v)}
   \sin\beta_u\nonumber\\
&-\alpha  \beta ^* \sqrt{(1-\eta ) \eta } \big(1-\cos\beta_u\big) e^{-i (\alpha_u-\alpha_v-\gamma_u)} \nonumber \\
&+\alpha^*  \beta \sqrt{(1-\eta ) \eta } \big(1+\cos\beta_u\big) e^{-i (\alpha_u-\alpha_v+\gamma_u)} \nonumber \\
&+\alpha  \beta ^* \sqrt{(1-\eta ) \eta } \big(1-\cos\beta_v\big) e^{i \gamma_v} \nonumber \\
&-\alpha^*  \beta \sqrt{(1-\eta ) \eta } \big(1+\cos\beta_v\big) e^{-i \gamma_v}.\label{P2}
\end{align}
It is obvious that the minimum $1/2$ of \eqref{purityV} is achieved when
\begin{equation}\label{eq:conds}
P_1=0 \ \ \text{and} \ \ P_2=0.
\end{equation}
The quantity \eqref{P1} vanishes when $\beta_u=0$ and $\beta_v=\pi$, which leads to $\theta=\pi$ and $\phi=-\pi$. With these values, the quantity \eqref{P2} vanishes with
\begin{align}
\zeta_{u}&:=\alpha_u+\gamma_u=\pi+\xi_v-2\theta_{\alpha\beta},
\end{align}
where
\begin{align}
\xi_v&:=\alpha_v-\gamma_v,\\
\theta_{\alpha \beta}&:=\theta_{\alpha}-\theta_{\beta}
\end{align}
and
\begin{subequations}\label{alphabeta}
\begin{eqnarray}
\alpha&=& r_{\alpha}e^{i\theta_{\alpha}} \\
\beta&=& r_{\beta}e^{i\theta_{\beta}},
\end{eqnarray}
\end{subequations}
with $r_{\beta}=\sqrt{1-r_{\alpha}^2}$.
Therefore, in this case having fixed $\theta=\pi$ and $\phi=-\pi$, the concurrence $C(\widetilde\rho'')$ remains function of four parameters,
i.e. $C(\eta,r_{\alpha},\theta_{\alpha \beta},\xi_v)$.

\begin{figure}[h]
 \centering
 \subfigure[\label{fig:conpurxip3thp2} $\theta_{\alpha\beta}=\frac{\pi}{2}$, $\xi_v=\frac{\pi}{3}$.]{\includegraphics[width=8cm,height=6cm,angle=0]{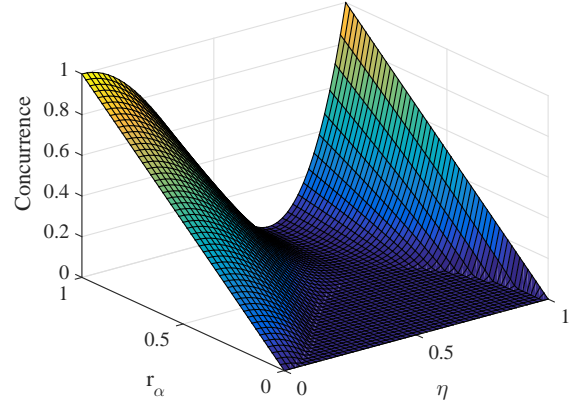}}
 \hspace{0.05\textwidth}
\subfigure[\label{fig:conpurxipthp0} $\theta_{\alpha\beta}=0$, $\xi_v=\pi$.]{\includegraphics[width=8cm,height=6cm,angle=0]{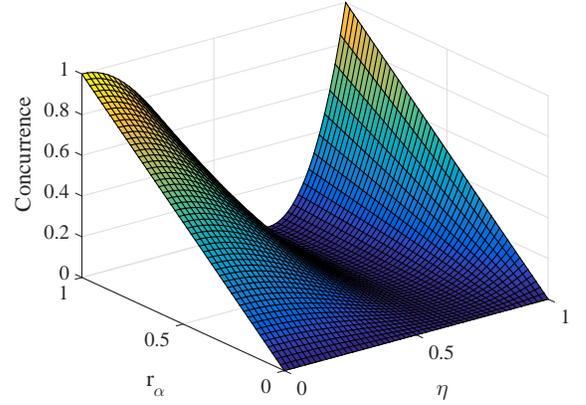}}
 	\caption{(Color online) Numerical results for concurrence under conditions \eqref{eq:conds}, i.e., $\beta_u=0$ and $\beta_v=\pi$ (or $\theta=\pi$ and $\phi=-\pi$), for some values of  $\theta_{\alpha\beta}$ and $\xi_v$.}
 	\label{fig:conpur}
 \end{figure}

 In order to find the maximum of concurrence over these four parameters and give a comparison with the concurrence of canonical Kraus operators \eqref{eq:Cfb}, we perform a numerical maximization over
$\eta$, $r_{\alpha}$, $\theta_{\alpha \beta}$ and $\xi_v$. This is done by choosing 11 values for $\eta$  and for $r_{\alpha}$ (varying them from 0 to 1 with step $0.1$),
as well as 61 values for $\theta_{\alpha \beta}$
and for $\xi_v$ (varying them from 0 to $2\pi$ with step $\pi/30$).
For any values of $\eta$, we obtain the maximum of concurrence over other $61^{2}\times 11$ points.
The numerical results show that the optimal concurrence is exactly the same as the one obtained in the canonical scenario, i.e. for $r_\alpha=1$.
Examples of numerical results are reported in Fig. \ref{fig:conpur}.

Taking into account the results of this and the previous Section (i.e. optimal feedback achieved for $\{\theta = \pi ,\phi = -\pi \}$) we end up with the following optimal local unitaries characterizing the feedback action in \eqref{Ufb}
\begin{equation}
\begin{aligned}\label{uvopt}
u&=\begin{pmatrix}
   e^{-i(\alpha_u+\gamma_u)/2} & 0 \\
   0 &  e^{i(\alpha_u+\gamma_u)/2}
   \end{pmatrix},
   \\ \\
v&=\begin{pmatrix}
   0 & -e^{-i(\alpha_v-\gamma_v)/2}  \\
   e^{i(\alpha_v-\gamma_v)/2} & 0
   \end{pmatrix},
\end{aligned}
\end{equation}
with arbitrary $\alpha_u+\gamma_u$ and $\alpha_v-\gamma_v$.


\section{Repeated feedback action}\label{sec:repeat}

Going back to Eq.\eqref{rhoppopt} we may observe that the matrix representation of $\rho^{\prime\prime}$ has
nonzero entries where also $\rho$ of Eq.\eqref{eq:initalstate} has.
Hence we may argue that the devised feedback action is optimal also starting from \eqref{rhoppopt}.

Then we repeat the analysis of Sections \ref{sec:FB} and \ref{sec:opt} starting from a state
 \begin{equation}
 	\begin{aligned}
 		\rho_q=\frac{1}{2}
 		&\begin{pmatrix}
 			1 & 0 & 0 & q \\
 			0 & 0 & 0 &0 \\
 			0 & 0 & 0 & 0 \\
 			q^* & 0 & 0 & 1
 		\end{pmatrix},
 	\end{aligned}
 	\label{inirhoq}
 \end{equation}
where $q$ is a generic complex number such that $|q|\le 1$.
In the computational basis $\mathfrak{B}$ and in the absence of feedback action, the state $\rho_q^\prime$ resulting from Eq.\eqref{dynnofbnew} reads
\begin{equation}
\begin{aligned}
\rho_q^\prime=\dfrac{1}{2}
&\begin{pmatrix}
\eta^2 & 0 & 0 & q\eta \\
0 & \eta(1-\eta) & 0 & 0 \\
0 & 0 & \eta(1-\eta) & 0 \\
q^*\eta & 0 & 0 & 2+\eta(\eta-2) \\
 \end{pmatrix}.
 \end{aligned}
 \label{dynnofbnewrhoq}
\end{equation}
Its subsystem purity is the same as Eq.(\ref{p1}) but its concurrence now depends on $|q|$. On the other hand, the matrix elements of $\rho_q''$ in the basis $\mathfrak{B}$ after applying \eqref{dynfbnew}, with \eqref{Ufb} and \eqref{eq:uv}, on the initial state (\ref{inirhoq}) result:
\begin{align}
\left[\rho_q''\right]_{{11}} &= \frac{1}{2} \bigg\{\left(1-\eta^2\right)\cos^2(\beta_v/2)+\eta^2 \sin^4(\beta_u/2)
\nonumber\\
&+2\eta(1-\eta)\cos ^2\left(\beta_v/2\right)\sin ^2\left(\beta_u/2\right) +\cos^2(\beta_u/2)  \nonumber\\
&+\eta\left( 1+ \cos\beta_u\right)\Re\left(q e^{-2i\gamma_u}\right) \sin ^2\left(\beta_u/2\right)\bigg\}, \label{eq:rq11} \\
\nonumber \\
\left[\rho_q''\right]_{12}&=\frac{1}{8} \bigg\{e^{-i \alpha_v}(1-\eta)^2 \sin \beta_v
\bigg[ \cos \beta_v+\frac{1-\eta\cos\beta_u}{1-\eta}\bigg] \nonumber \\
  &+e^{-i \alpha_u}\sin\beta_u\Big[(1-\eta)(1-\eta \cos \beta_v) +(1+\eta ^2)\cos \beta_u  \nonumber\\
  &-2\eta\cos\beta_u\Re\left(q e^{-2i\gamma_u}\right) -2i\eta\Im\left(q e^{-2i\gamma_u}\right) \Big] \bigg\}, \label{eq:rq12}
\end{align}

\begin{align}
\left[\rho_q''\right]_{13}&= \left[\rho_q''\right]_{12}, \\ \nonumber\\
\left[\rho_q''\right]_{14}&=\frac{1}{8}e^{-2 i (\alpha_u+\gamma_u)}\bigg\{ \eta q(1+ \cos \beta_u)^2 \nonumber \\
&+ 4e^{2 i\gamma_u}\left[q^*\eta e^{2i\gamma_u} +(1+\eta^2)\cos^2(\beta_u/2)\right] \sin ^2\left(\beta_u/2\right) \nonumber \\
&+ 2e^{2 i (\alpha_u-\alpha_v+\gamma_u)}(1-\eta)^2(1+\cos \beta_v) \sin ^2\left(\beta_v/2\right) \nonumber \\
&-2e^{ i (\alpha_u-\alpha_v+2\gamma_u)}\eta(1-\eta)\sin \beta_u\sin \beta_v
\bigg\},
\end{align}

\begin{align}
\left[\rho_q''\right]_{22}&= \frac{1}{8} \bigg\{4 (1-\eta) \eta  \cos ^2\left(\beta_u/2\right)
   \cos ^2\left(\beta_v/2\right) \nonumber\\
 &+ \left(1+\eta ^2-2 \eta  \Re\left(q e^{-2i\gamma_u}\right)\right)\sin ^2\beta_u  \nonumber \\
 &+2 (1-\eta) \big[1-\eta
    \cos\beta_u+(1-\eta) \cos \beta_v\big]\sin ^2\left(\beta_v/2\right) \bigg\}, \label{eq:rq22}
\\ \nonumber \\
\left[\rho_q''\right]_{23}&=\frac{1}{8}\bigg\{\left(1+\eta ^2-2 \eta  \Re\left(q e^{-2i\gamma_u}\right) \right)\sin ^2\beta_u\nonumber\\
&-(1-\eta)\sin\beta_v \Big[2\eta\cos(\alpha_u-\alpha_v)\sin \beta_u \nonumber\\
&-(1-\eta)\sin\beta_v \Big]\bigg\}, \label{eq:rq23}
\end{align}

\begin{align}
\left[\rho_q''\right]_{24}&=\frac{1}{8} \bigg\{
e^{-i \alpha_u}\sin\beta_u\Big[(1-\eta)(1+\eta \cos \beta_v) \nonumber\\
&-\left(1+\eta^2
-2\eta\Re\left(q e^{-2i\gamma_u}\right)\right)\cos\beta_u +2i\eta\Im\left(q e^{-2i\gamma_u}\right) \Big] \nonumber\\
&+ e^{-i \alpha_v}(1-\eta) \big[1+\eta\cos\beta_u-(1-\eta) \cos \beta_v\big]\sin \beta_v
  \bigg\}, \label{eq:rq24}\\
  \nonumber\\
\left[\rho_q''\right]_{33}&= \left[\rho_q''\right]_{22},
\end{align}

\begin{align}
\left[\rho_q''\right]_{34}&= \left[\rho_q''\right]_{24}, \\ \nonumber\\
\left[\rho_q''\right]_{44}&=\frac{1}{8} \bigg\{ \eta^2\left( 1+ \cos\beta_u\right)^2+4(1-\eta)^2\sin^4\left(\beta_v/2\right) \nonumber\\
&+4\sin^4\left(\beta_u/2\right)+8 (1-\eta) \eta  \cos ^2\left(\beta_u/2\right)
    \sin ^2\left(\beta_v/2\right) \nonumber\\
&+4\eta(1-\cos^2\beta_u)\Re\left(q e^{-2i\gamma_u}\right)\bigg\}.
 \label{eq:rq44}
\end{align}
For the state $\rho_q''$, the subsystem purity turns out to be the same of \eqref{p2}, i.e. not depending
on $q$. This leads us to conclude that also for $|q|<1$ the optimal feedback is achieved
by $\{\theta = \pi ,\phi = -\pi \}$ and hence \eqref{uvopt}. With this, the state after feedback action reads, 
in the basis $\mathfrak{B}$,
 \begin{equation}
\begin{aligned}
\rho_q''=\frac{1}{2}
&\begin{pmatrix}
1 & 0 & 0 & q \eta e^{-2 i (\alpha_u+\gamma_u)} \\
0 & 0 & 0 &0 \\
0 & 0 & 0 & 0 \\
q^* \eta e^{2 i (\alpha_u+\gamma_u)} & 0 & 0 & 1
 \end{pmatrix}.
 \end{aligned}
 \label{rhoppoptq}
\end{equation}
Its concurrence results
\begin{equation}\label{eq:Cfbq}
C\left(\rho_q''\right)=|q| \, \eta.
\end{equation}
The optimality of this result is confirmed by numerical investigations over non-canonical Kraus decompositions \eqref{eq:tildeKj}. Similarly to Sec. \ref{sec:opt} we have maximized the concurrence $C(\rho_q'')$ over parameters $r_{\alpha}$, $\theta_{\alpha \beta}$ and $\xi_v$, this time for each pair of values of $\eta$ and $q$. This has been done by choosing 11 values for $\eta$, for $|q|$ and for $r_\alpha$ (varying them between 0 and 1 with step $0.1$), as well as 61 values for $\theta_{\alpha \beta}$ and for $\xi_v$ (varying them from 0 to $2\pi$ with step $\pi/30$).
For any pair of $\eta$ and $|q|$ the maximum concurrence has been obtained over other $61^2 \times 11$ points. 
The numerical results show that the optimal concurrence is exactly \eqref{eq:Cfbq}, i.e. the one obtained in the canonical scenario ($r_\alpha=1$).

\begin{figure}
 \centering
	\includegraphics[width=8cm,height=6cm,angle=0]{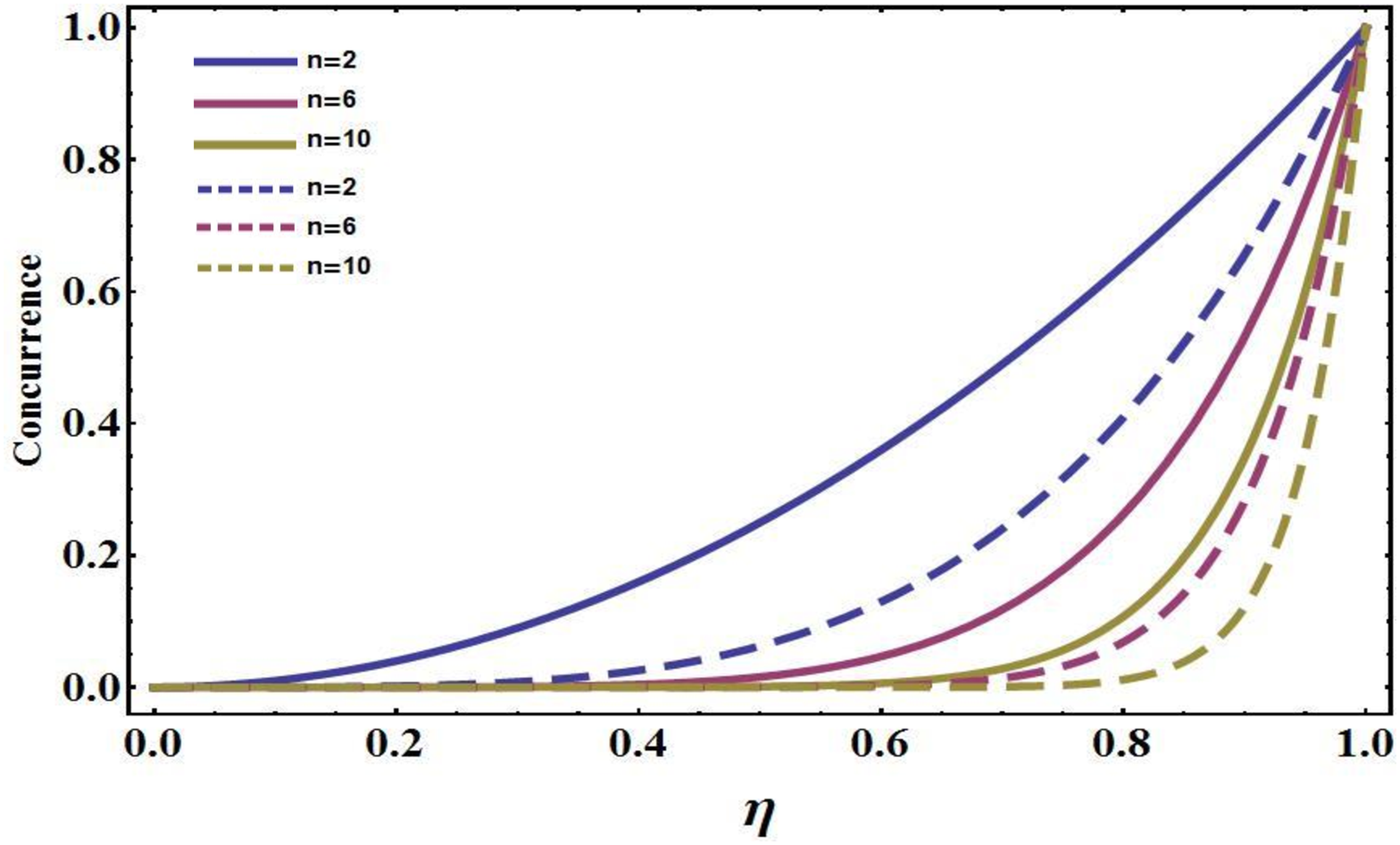}
	\caption{Concurrence versus $\eta$ for different number $n$ of applications of the amplitude damping map, without feedback action (dashed line) and with feedback action (solid line).}
	\label{num_feed_n}
\end{figure}

Thanks to the above results, we can consider
repeated applications of the map without feedback, giving
 \begin{equation}
 \begin{aligned}
 \rho^{\prime(n)}=\frac{\eta^n}{2}
 &\begin{pmatrix}
 \eta^{n} & 0 & 0 & 1  \\
 0 & 1-\eta^n & 0 &0 \\
 0 & 0 & 1-\eta^n & 0 \\
 1 & 0 & 0 & \frac{2}{\eta^n}-(2-\eta^n)
 \end{pmatrix},
 \end{aligned}
 \label{mapn}
 \end{equation}
where $n$ is the number of map's applications, as well as
repeated applications of the map with feedback giving
\begin{equation}
 	\begin{aligned}
 		\rho^{\prime\prime(n)}=\frac{1}{2}
 		&\begin{pmatrix}
 			1 & 0 & 0 & \eta^n e^{-2 n i (\alpha_u+\gamma_u)} \\
 			0 & 0 & 0 &0 \\
 			0 & 0 & 0 & 0 \\
 			\eta^n e^{2 n i (\alpha_u+\gamma_u)} & 0 & 0 & 1
 		\end{pmatrix}.
 	\end{aligned}
 	\label{fbn}
 \end{equation}
The corresponding concurrences
\begin{equation}
\begin{aligned}
C\left(\rho^{\prime (n)}\right)=&\frac{\eta^n}{2} 
\bigg[ \sqrt{\left(\eta ^n-2\right) \eta ^n+3+2 \sqrt{\left(\eta ^n-2\right) \eta ^n+2}}\\
&-\sqrt{\left(\eta ^n-2\right) \eta ^n+3-2 \sqrt{\left(\eta ^n-2\right) \eta ^n+2}}\\
&-2 (1-\eta ^n) \bigg],
\end{aligned}
\end{equation}   
and 
\begin{equation}
 C\left(\rho^{\prime\prime (n)}\right)=\eta ^n,
\end{equation} 
are reported in Fig.\ref{num_feed_n}. There we can see that the advantage of feedback tends to persist only at sufficiently high values of $\eta$,
by increasing $n$.

\section{Conclusion}\label{sec:conclu}

In conclusion, we have addressed the problem of correcting errors intervening in two-qubit dissipating into their own environments by resorting to local feedback actions with the aim of preserving as much as possible the initial amount of entanglement. Optimal control is found by first gaining insights from the subsystem purity and then by numerical analysis on the concurrence.
This is tantamount to a double optimization, on the actuation and on the measurement
precesses. The results are obtained for single shot.
The results, although obtained with the help of numerics, are analytically clear and can be summarized
by Eqs. \eqref{dynfbnew} and \eqref{Ufb} with \eqref{uvopt}.

Our results could be helpful in designing experiments where entanglement control is required, particularly in settings like cavity QED \cite{Maunz2004}, superconducting qubits \cite{You2011}, optomechanical systems \cite{Aspelmeyer2014}.

It remains open the problem of steering the system towards a desired target (entangled) state; to this end we need to consider repeated map's applications for which we paved the way in Section \ref{sec:repeat}.
The feedback strategy employed along this line is in the same spirit of \emph{direct feedback} \cite{WM93},
in that it does not involve processing the information obtained from the system in order to estimate its state.
On the other hand in the context of repeated map's applications, and particularly in the continuous time analysis of the problem, optimization of feedback action should also involve Bayesian (state estimation based) strategies
and an extension to two qubits of the analysis for single qubit control performed in Ref.\cite{WMW2002} would be
very welcome.


\end{document}